\documentclass[12pt]{iopart}
\usepackage[utf8]{inputenc}
\usepackage{graphicx}
\usepackage{graphics}
\usepackage{dcolumn}
\usepackage{bm}
\usepackage{float}

\usepackage[labelsep=space]{caption}

\begin{document}
\title{Study of Nuclear Effects in the Computation of the $0\nu\beta\beta$ Decay Matrix Elements}
\author{Andrei Neacsu$^1$, Sabin Stoica$^2$}
\address{Horia Hulubei Foundation, P.O. Box MG-12 and \\Horia Hulubei National Institute for Physics and Nuclear Engineering, \\P.O. Box MG-6, Magurele-Bucharest 077125, Romania}
\ead{$^1$nandrei@theory.nipne.ro, $^2$stoica@theory.nipne.ro}
\begin{abstract}

We analyze the effects that different nuclear structure approximations associated with the short range correlations (SRC), finite nucleon size (FNS), higher order terms in the nucleon currents (HOC) and with some nuclear input parameters, have on the values of the nuclear matrix elements (NMEs) for the neutrinoless double beta ($0\nu\beta\beta$) decay.
The calculations are performed with a new Shell Model(ShM) code which allows a fast computation of the two-body matrix elements of the transition operators. The treatment of SRC, FNS and HOC and the use of quenched or unquenched values for the axial vector coupling constant produces the most important effects on the NMEs values. Equivalent effects of some of these approximations are also possible, which may lead (accidentally) to close final results.  We found that the cumulative effect of all these nuclear ingredients on the calculated nuclear matrix elements NMEs is significant. Since the NMEs values are often obtained with different approximations and/or with different input parameters, a convergent view point on their inclusion/neglecting and an uniformization of the calculations are needed, in order to enter in an era of precision concerning the computation of the NMEs for double beta decay.  
\end{abstract}
\maketitle

\eqnobysec
\section{Introduction}
Neutrinoless double beta ($0 \nu \beta \beta$) decay is a beyond Standard Model (SM) process by which an even-even nucleus transforms into another even-even nucleus with the emission of two electrons/positrons but no antineutrinos/neutrinos in the final states. 
Its discovery would clarify the question about the lepton number conservation, decide on the neutrinos character (are they distinguished from their antiparticles?) and give a hint on the scale of their absolute masses. 
The importance of these fundamental issues has led to extended theoretical and experimental investigations of this process. 
The reader can find an up-to-date information on these studies from several recent reviews \cite{AEE08}-\cite{ROJ12}, which also contain therein a comprehensive list of references in the domain.

For correctly predicting neutrinoless double beta decay (DBD) lifetimes and getting information about the neutrino properties, a required ingredient is the accurate value of the NMEs. There are still important differences between the results reported in literature, hence their precise calculation continues to be an important challenge in the study of the DBD process. 
The NMEs are currently computed by several methods, the proton-neutron Quasi Random Phase Approximation (pnQRPA) \cite{ROD07}-\cite{SK01}, Interacting Shell Model(ISM) \cite{Cau95}-\cite{HS10}, Interacting Boson Approximation (IBA) \cite{BI09}-\cite{iba-prc2013}, Projected Hartree Fock Bogoliubov (PHFB) \cite{RAH10} and Energy Density Functional Method \cite{RMP10} (EDF) being the present most employed methods. 
Each of them has its own advantages and drawbacks, and these have been largely discussed in the literature (see for example \cite{VOG12}-\cite{FAE12}).

The differences between the calculated values of the NMEs may come, on the one hand, from the specific assumptions of the methods employed, and, on the other hand, from the different nuclear ingredients and parameters involved in their calculation. 
Particularly, significant differences can come from the manner of inclusion of SRC and FNS effects, and of taking into account HOC including the tensor term. 

Also, there is still an open question on the values of some nuclear input parameters needed in calculations, such as the axial coupling constant $g_A$ (a quenched or unquenched value), or the value of the average excited intermediate energy $\left< E \right>$ used in the closure approximation. 
Other constants whose values may differ in different calculations are $r_0$ from the expression of the mean nuclear radius $R=r_0A^{1/3}$ and the cutoff values used in the nucleon form factors, $\Lambda_V$, $\Lambda_A$, also induce changes in the calculated NMEs final results that can be quantized as well.
Finally, an additional source of uncertainty is also the nucleon-nucleon (NN) effective interaction employed in calculations. 
Until now, the SRC, FNS and HOC effects have been the most studied in QRPA calculations  \cite{VOG12}-\cite{FAE12}, \cite{KOR07} and \cite{SIM08}-\cite{sim-09}, and to a less extent in ISM \cite{npa818} and IBA2 \cite{iba-prc2013} calculations.
At present, there is not yet a consensus among theorists concerning their results, which leads to difficulties in comparing the NMEs from literature due to the use of different nuclear ingredients and/or parameters in calculations. 
However, their cumulated effect on the NMEs values is important, and hence a detailed analysis performed within all of the methods is needed for a better comparison of the different results and for achieving a consensus on their use in calculations.
 
The NMEs calculations are performed with a new ShM code. In a previous paper \cite{NSH12}, we shortly presented a new, improved (fast, efficient) ShM code, which reduces substantially the computing time of calculation of the two-body matrix elements (TBMEs) of the transition operators for the  $0\nu\beta\beta$ decay. 
It incorporates all the relevant nuclear effects. The main improvement of this code comes from a rearrangement in the TBMEs expression, which allows us to analytically perform the radial integrals (the integrals over the coordinates space), when harmonic oscillator $\left(HO\right)$ single particle w.f. are used. 
Therefore, only the integration over the momentum remains to be performed numerically. We found this code to about 30 times faster than our previous code \cite{HS10}, for the same calculations, and it can be of much help for investigating the quenching of the Gammow-Teller (GT) operator in ShM calculations of the NMEs. 
For the $2 \nu \beta \beta$ decay mode, the GT operator needs to be quenched, to better describe the experimental data for beta decays and charge-exchange reactions, while for the $0 \nu \beta \beta$ there is no evidence for it. Studies on this issue have been done in \cite{EH09}-\cite{medex11-mh}.
However, to investigate this effect in other real, more complicated cases, one needs to know the TBMEs of the bare transition operator in larger model spaces (e.g. composed by 8 to 12 major harmonic oscillator shells). 
When the HOC are also included, the calculations may face severe computational problems due to the very long CPU times required, and hence fast numerical codes for calculating the TBMEs are needed.

The calculation of the NMEs for the $0\nu \beta \beta$ decay of three isotopes, i.e. $^{48}Ca$, $^{76}Ge$, $^{82}Se$, was performed with an improved version of our code from \cite{NSH12}, including the tensor part of the neutrino potential, thus, completing the inclusion of all HOC terms which appear in the expression of the neutrino potential. 
We analyze the effects of different nuclear ingredients such as SRC, FNS, HOC, the axial vector constant and the average excited energy used in the closure approximation, on the  NMEs for $0\nu\beta\beta$ decay. 
We gradually introduce the nuclear structure effects, in order to see their individual and cumulative contribution to the final value of the NMEs. Also, we perform the calculations with different values of $g_A$ (quenched and unquenched), $\left< E\right>$, $r_0$ and cutoff parameters, to observe the errors associated with these input parameters.
To provide the reader with more detailed information, we also report the values of all the $M^{0\nu}$ components (magnitude and sign), i.e. GT, F and tensor components. 

The paper is organized as follows. In Section 2 we describe in more detail our new ShM code. In Section 3 we present our values for the NMEs, discuss the influence of the nuclear ingredients/parameters mentioned above on the results, and compare our results with other similar ones from literature.  
The last section is devoted to some final remarks and conclusions. Finally, in Annex, we give the analytical formulae of several quantities used in our ShM code.

\section{Calculation of the nuclear matrix elements}
In the ``standard'' scenario, when the $0\nu\beta\beta$ decay process occurs by exchange of light Majorana neutrinos between two nucleons inside the nucleus, and in the presence of left-handed weak interactions, the lifetime expression can be written, in a good approximation, as a product of three factors:
\begin{equation}
\left( T^{0\nu}_{1/2} \right)^{-1}=G^{0\nu}(Q_{\beta\beta}, Z)\mid M^{0\nu}\mid^2 \left( \frac{ \left< m_{\nu} \right> }{m_e} \right)^2 \ ,
\end{equation}
$G^{0\nu}$ is the phase space factor for this decay mode, depending on the energy decay $Q_{\beta\beta}$ and nuclear charge $Z$, $\left<m_\nu\right>$ is the effective neutrino mass parameter depending on the first row elements $U_{ei} (i=1,2.3)$ of the neutrino mixing matrix, $m_e$ is the electron mass and $M^{0\nu}$ are the NMEs depending on the nuclear structure of the nuclei involved in the decay.
The expression of the NMEs can be written, in general, as a sum of three components:
\begin{equation}
 M^{0 \nu}=M^{0 \nu}_{GT}-\left( \frac{g_V}{g_A} \right)^2  M^{0 \nu}_F + M^{0 \nu}_T \ ,
 \label{nme}
\end{equation}
where $M^{0 \nu}_{GT}$, $M^{0 \nu}_F$ and $M^{0 \nu}_T$ are the Gamow-Teller ($GT$), Fermi ($F$) and tensor ($T$) components, respectively. These are defined as follows:
\begin{equation}
M_\alpha^{0\nu} = \sum_{m,n} \left< 0^+_f\| \tau_{-m} \tau_{-n}O^\alpha_{mn}\|0^+_i \right> \ ,
\end{equation}
where $O^\alpha_{mn}$ are transition operators ($\alpha=GT,F,T$) and the summation is over all the nucleon states.

Due to the two-body nature of the transition operator, the NMEs can be also expressed as a sum of products of two-body transition densities (TBTDs)  and matrix elements of the two-body transition operators for two-particle states, shortly, two-body matrix elements (TBMEs),
\begin{equation}
\fl M_\alpha^{0\nu}= \sum_{j_p j_{p^\prime}j_n j_{n^\prime}J_\pi}TBTD \left( j_p j_{p^\prime},j_n j_{n^\prime};J_\pi \right) \left< j_p j_{p^\prime};J_\pi \| \tau_{-1} \tau_{-2}O^\alpha_{12} \| j_n j_{n^\prime};S_\alpha J_\pi \right> ,
\end{equation}
where $|jj';J^\pi>$ represent the antisymmetrized two-particle states. Since the NN effective interaction can be treated by means of a central (single-particle) potential, the NMEs can be conveniently calculated using Moshinsky's transformations between the relative and Center of mass (CM) coordinates and the proper use of nuclear states in different coupling notations.
Correspondingly, the two-body transition operators $O^{\alpha}_{12}$ can be expressed in a factorized form as:
\begin{eqnarray}
O^{\alpha}_{12} = N_{\alpha}  S_{\alpha}^{(k)} \cdot  \left[R_{\alpha}^{(k_r)}\times C_{\alpha}^{(k_c)}\right]^{(k)}
\end{eqnarray}
where $N_\alpha$ is a numerical factor including the coupling constants, and  $S_\alpha$, $R_\alpha$ and $C_\alpha$ are operators acting on the spin, relative and CM wave functions of two-particle states, respectively.
Thus, the calculation of the matrix elements of these operators can be decomposed into products of reduced matrix elements within the two subspaces \cite{HS10}. The expressions of the two-body transition operators are:
\begin{equation}
\fl O_{12}^{GT} = \sigma _1 \cdot \sigma _2 H(r) \ , \ \ \ \ \
O_{12}^{F} = H(r) \ , \ \ \ \ \
O_{12}^{T} = \sqrt{\frac{2}{3}} \left [ \sigma _1 \times \sigma _2 \right ]^2 \cdot \frac{r}{R} H(r) C^{(2)}(\hat r) \ 
\end{equation}
The most difficult is the computation of the radial part of the two-body transition operators, which contains the neutrino potential.  We will refer to this issue in more detail.
The neutrino potential depends weakly on the intermediate states, and is defined by integrals of momentum carried by the virtual neutrino exchanged between the two nucleons \cite{sim-09}
\begin{equation}
\fl H_{\alpha} (r) = \frac{2R}{\pi}  \int^\infty_0 j_i (qr) \frac{h_{\alpha}(q)}{\omega} \frac{1}{\omega + \left<E\right>}q^2 dq \equiv \int^\infty_0 j_i (qr) V_{\alpha}(q) q^2 dq \ ,
\label{n_potential}
\end{equation}
where $R=r_0 A^{1/3}$ fm, with $r_0=1.2fm$, $\omega = \sqrt{q^2+m_\nu^2}$ is the neutrino energy and $j_i(qr)$ is the spherical Bessel function (i = 0, 0 and 2 for GT, F, and T components, respectively).
We use the closure approximation in our calculations, and $\left<E\right>$ represents the average excitation energy of the states in the intermediate odd-odd nucleus, that contribute to the decay. The expressions of $h_{\alpha} (\alpha = F, GT$ and $T)$ are
\begin{equation}
\fl h_F = G_V^2(q^2) \ ,
\label{hf-hoc}
\end{equation}
\begin{equation}
\fl h_{GT}(q^2) = \frac {G^{2}_A(q^2)}{g^{2}_A} \left [ 1- \frac{2}{3}\frac{q^2}{q^2+m^2_\pi} + \frac{1}{3}\left( \frac{q^2}{q^2+m^2_\pi} \right )^2 \right ]+ \frac{2}{3} \frac {G^{2}_M(q^2)}{g^{2}_A}\frac{q^2}{4m^2_p} \ ,
\label{hgt-hoc}
\end{equation}
and
\begin{equation}
\fl h_{T}(q^2) = \frac {G^{2}_A(q^2)}{g^{2}_A} \left [\frac{2}{3}\frac{q^2}{q^2+m^2_\pi} - \frac{1}{3}\left( \frac{q^2}{q^2+m^2_\pi} \right )^2 \right ]+ \frac{1}{3} \frac {G^{2}_M(q^2)}{g^{2}_A}\frac{q^2}{4m^2_p} \ ,
\label{ht-hoc}
\end{equation}
where $m_\pi$ is the pion mass, $m_p$ is the proton mass and
 \begin{equation}
G_M(q^2) = (\mu_p - \mu_n)G_V(q^2),
\end{equation}
with $(\mu_p - \mu_n)=4.71$. \\
The expression (\ref{hf-hoc}) includes FNS effects, while the expressions (\ref{hgt-hoc}) and (\ref{ht-hoc}) include both FNS and HOC effects.
The $G_V$ and $G_A$ form factors, which takes into account the finite size of the nucleons effect, are:
\begin{equation}
G_A \left(q^2 \right) = g_A \left( \frac{\Lambda^2_A}{\Lambda^2_A+q^2} \right)^2, \ G_V \left( q^2 \right) = g_V \left( \frac{\Lambda^2_V}{\Lambda^2_V+q^2} \right)^2
\label{formfactors}
\end{equation}
For the vector and axial coupling constants the majority of calculations take $g_V = 1$ and the unquenched value $g_A = 1.25$, while the values of the vector and axial vectors form factors are $\Lambda_V=850 MeV$ and $\Lambda_A=1086 MeV$ \cite{AEE08}, respectively.

For computing the radial matrix elements $\left<nl|H_{\alpha}|n'l'\right>$ we use
the $HO$ w. f.  $\psi_{nl}(lr)$ and $\psi_{n^\prime l^\prime}(r)$ corrected by a factor $[1 + f(r)]$, which takes into account the SRC  induced by the nuclear interaction:
\begin{equation}
\psi_{nl}(r) \rightarrow \left[ 1+f(r) \right] \psi_{nl}(r) \ 
\end{equation}
For the correlation function we take the functional form
\begin{eqnarray}
f(r) = - c \cdot e^{-ar^2} \left( 1-br^2 \right) \ ,
\label{src}
\end{eqnarray}
where $a$, $b$ and $c$ are constants which have particular values for different parameterizations \cite{sim-09}, as it will be discussed in the next section. For $c=1$, we get the Jastrow prescription \cite{jastrow55} of inclusion of the SRC effects, and in this case, the $a$ and $b$ constants are given by the Miller-Spencer (MS) parameterization \cite{jastrow}.

Including HOC and FNS effects, the radial matrix elements of the neutrino potentials become:
\begin{equation}
\fl \left\langle nl \mid H_\alpha(r) \mid n^\prime l^\prime \right\rangle \ = \ \int^\infty_0 r^2 dr \psi_{nl}(r) \psi_{n^\prime l^\prime}(r)\left[ 1+f(r) \right]^2 \times \int^\infty_0 q^2 dq V_{\alpha} (q) j_n (qr) \ ,
\label{H-two-integrals}
\end{equation}
where $\nu$ is the oscillator constant.

As one can see, if all the nuclear effects are included, the calculation of the radial integrals (\ref{H-two-integrals}) requires the numerical computation of two integrals, one over the coordinate space and the other over the momentum space. 
However, one can reduce the computation to only one integral by rearranging the expression of the radial integral in coordinate space as a sum of terms with the same power of $r$ \cite{NSH12}. By this procedure, which is described in more detail in Annex, 
the expression of the radial matrix element for the neutrino potential can be written as follows,
\begin{equation}
\left\langle nl\mid H_{\alpha}(r)\mid n^{\prime} l^{\prime} \right\rangle = \sum_{s=0}^{n+n^{\prime}} A_{l+l^{\prime}+2s}(nl,n^{\prime}l^{\prime}) \mathcal{K}_{\alpha}(m)
\label{finaleq}
\end{equation}
where $\mathcal{K}_{\alpha}(m)$ is a sum of six integrals over momentum.
The expressions for $A_{l+l^{\prime}+2s}(nl,n^{\prime}l^{\prime})$ and $\mathcal{K}_{\alpha}(m)$ are given in Annex.

\section{Numerical results and discussions}
We developed a new code for computing the TBMEs necessary for the ShM calculations of the NMEs involved in $0\nu\beta\beta$ decays, based on the formalism described in the previous section. Our code can include, in a flexible manner, different nuclear effects, such as FNS, HOC and SRC introduced either by the Jastrow \cite{jastrow} prescription with Miller-Spencer (MS) parameterization or by coupled cluster method (CCM) with Argonne V18 and CD-Bonn \cite{sim-09}, \cite{ccm},\cite{ccm1} NN forces. 
The SRC parameters entering (\ref{src}) are the same as in table II of \cite{HS10}.
 The two-body transition densities needed to calculate the $M^{0\nu}$ matrix element, are computed with the code ANTOINE \cite{ANT} and using the method described in \cite{HS10}. For $^{48}Ca$ we use two different NN effective interactions GXPF1A \cite{gxpf1a} and KB3G \cite{KB68} in the full $pf$ shell as model space, while for $^{76}Ge$ and $^{82}Se$ we use JUN-45 \cite{jun45} effective interactions in the $1p_{3/2},0f_{5/2},1p_{1/2},0g_{9/2}$ valence space ($jj44$). 
 We perform the calculations within the closure approximation, with the value of the average energy $\left< E \right>$ given by the formula $\left< E \right> =1.12A^{1/2} MeV$. For the other nuclear input parameters we used the values $r_0=1.2fm$ and $\Lambda_V=850 MeV$, $\Lambda_A=1086 MeV$. 
Our results for the total NMEs values, $M^{0\nu}_{total}$, are presented in table \ref{tab-effecs}, for the three isotopes studied, i.e. $^{48}Ca$, $^{76}Ge$ and $^{Se}$. We included gradually the nuclear effects in calculation in order to have a complete view of their influence on the NMEs. 
\begin{table}
\begin{center}
\caption[E=auto]{. The NMEs obtained with inclusion of different nuclear effects. ''b'' denotes the value obtained without any effect included, while ``F'', H'' ``S'' and ``total'' indices denote the $M^{0\nu}$ values obtained when FNS, HOC, SRC and all effects, are, respectively, included. The set of the three values from the columns with SRC effects included refers to the particular prescriptions: (a)=Jastrow with MS parameterization, (b)=CCM-AV18 and (c)=CCM-CD-Bonn type. The calculations are performed with $g_A$=1.25, $r_0=1.2fm$, $\Lambda_V=850 MeV$, $\Lambda_A=1086 MeV$.}
\begin{tabular}{@{}crrrrrrcr@{}}   \hline
\label{tab-effecs}
 &$M_{b}$&$M_{b+F}$&$M_{b+H}$&$M_{b+F+H}$&$M_{b+S}$&$M_{b+S+F}$&$M_{b+S+H}$&$M^{0\nu}_{total}$  \\ \hline
			       & & & & &(a)-0.731&-0.680&-0.542&-0.508\\
  $^{48}Ca$&-1.166&-0.959&-0.923&-0.773&(b)-1.023&-0.930&-0.800&-0.733\\
			       & & & & &(c)-1.153&-1.008&-0.914&-0.809\\ \hline

			     & & & & &(a)\ 0.856&0.798&0.670&0.628\\
  $^{48}Ca^*$&1.351&1.116&1.102&0.928&(b)\ 1.188&1.082&0.962&0.884\\
			     & & & & &(c)\ 1.337&1.171&1.092&0.969\\ \hline

			   & & & & &(a)\ 3.025&2.889&2.499&2.378\\
  $^{76}Ge$&4.168&3.615&3.497&3.066&(b)\  3.807&3.557&3.187&2.979\\
			   & & & & &(c)\  4.153&3.762&3.489&3.177\\ \hline

				&& & & &(a)-2.779&-2.665&-2.275&-2.176\\
  $^{82}Se$&-3.779&-3.305&-3.140&-2.780&(b)-3.467&-3.256&-2.876&-2.703\\
				&& & & &(c)-3.770&-3.438&-3.137&-2.878\\ \hline
  \end{tabular}
  \end{center}
\end{table}
The $M^{0\nu}$ values, where different nuclear effects are included, are shown separately, in order to appreciate their individual and/or combined contributions. 
The ``b''(bare) notation means the $M^{0\nu}$ value obtained with any nuclear effect included, while ``F'', H'' ``S'' and ``total'' indices denote the $M^{0\nu}$ values obtained when FNS, HOC, SRC and all effects, respectively, are included. 
The set of the three values from the columns with SRC effects included refers to the particular prescriptions: (a)=Jastrow with MS parameterization (J-MS), (b)= CCM-AV18 and (c)=CCM-CDBonn type. Generally, the features associated with these nuclear effects, previously observed and discussed in literature, are confirmed by our calculations and will be discussed in the following. We remark first that the inclusion of these nuclear effects diminishes gradually the ``bare'' values of the NMEs. The cumulative effect of all the nuclear ingredients on $M^{0\nu}_b$, presented in table \ref{tab-effecs} is between 43-56\% for SRC with J-MS prescription, while for the other two SRC prescriptions, CCM-AV18 and CCM-CDBonn, the effect is milder, 24-31\%. Thus, after 2007, when SRC were introduced by softer prescriptions, UCOM \cite{suhonen-ucom} and CCM \cite{sim-09}, there was a significant increase of the values of $M^{0\nu}_{total}$ in literature. In our present calculations this increase is of about 36\% for $^{48}Ca$ and 24\% for $^{76}Ge$ and $^{82}Se$.  
It is worth to mention that the inclusion of only SRC with J-MS produces almost the same effect on the $M^{0\nu}_b$ value as that when all nuclear effects are included, but with SRC introduced by ``softer'' prescriptions, giving NMEs values which are close to the $M^{0\nu}_{total}$ ones within 4\% for the CCM-AV18 and 10\% for CCM-CDBonn parameterizations. This is an example of possible equivalent effects of these different nuclear ingredients taken into account in calculations.  
However, this agreement is rather an accidental one, based on a too strong cut of the radial w.f. at short distances produced by the SRC with J-MS, and the correct prescription for calculation, agreed upon in the literature, is to properly include all the nuclear effects by softer SRC prescriptions, such as CCM or UCOM. 
Further, we mainly refer to the influence of the nuclear effects when SRC with CCM prescriptions are included. The effects of SRC and FNS are not additive, but their simultaneous inclusion compensates somehow their global(total) effect on $M^{0\nu}_b$. 
The inclusion of HOC is important, The neglecting of these corrections results in larger values of $M^{0\nu}_{total}$ with $\sim 20\%$ for $^{48}Ca$ and $\sim 16\%$ for $^{76}Ge$ and $^{82}Se$. Besides these nuclear effects, one also remarks that the use of different NN interactions, in the case of $^{48}Ca$, can induce differences of about $17\%$ between NMEs values obtained within the same approximation. Thus, one can expect that the use of different NN interactions can also produce significant differences between the calculated NMEs, as well.

In table \ref{tab-ga} we show the values of each component (GT, F and T) of $M^{0\nu}_{total}$ (according to (\ref{nme})), in order to see their individual contributions (magnitude and sign), and provide the reader with more information on the calculations. 
\begin{table}[b]
\begin{center}
\caption[results]{. Values of the GT, Fermi = F, and tensor = T components of the total NMEs, $M^{0\nu}_{total}$. $^{48}$Ca NME is computed with the GXPF1A\cite{gxpf1a} and $^*$KB3G\cite{KB68} effective NN interactions in the $pf$ major shell, while $^{76}$Ge and $^{82}$Se NMEs are computed with the JUN45\cite{jun45} effective NN interaction. HOC, FNS and CD-BONN SRC effects are included. The calculations are performed with three values of $g_A$, which are displayed.}
\begin{tabular}{@{}clrrlr@{}}  \hline
\label{tab-ga}
&$g_A$&$M^{0\nu}_{total}$&$M^{0\nu}_{GT}$&$-{\left( \frac{gV}{gA} \right) }^2 \ M^{0\nu}_{F}$&$M^{0\nu}_{T}$ \\ \hline
	   &1.00&-0.929&-0.789&\ -1.00\ \ \ 0.223&0.083\\
  $^{48}Ca$&1.25&-0.809&-0.740&\ -0.64\ \ \ 0.223&0.074\\
	  &1.275&-0.800&-0.737&\ -0.62\ \ \ 0.223&0.074\\ \hline

	     &1.00&1.106&0.925&\ -1.00\ \ -0.244&-0.063\\
  $^{48}Ca^*$&1.25&0.969&0.870&\ -0.64\ \ -0.244&-0.057\\
	    &1.275&0.960&0.866&\ -0.62\ \ -0.244&-0.056\\ \hline

	   &1.00&3.540&3.030&\ -1.00\ \ -0.640&-0.130\\
  $^{76}Ge$&1.25&3.177&2.897&\ -0.64\ \ -0.640&-0.130\\
	  &1.275&3.152&2.890&\ -0.62\ \ -0.640&-0.130\\ \hline

	   &1.00&-3.361&-2.877&\ -1.00 \ \  0.607&0.123 \\
  $^{82}Se$&1.25&-2.878&-2.752&\ -0.64 \ \  0.607&0.123 \\
	  &1.275&-2.850&-2.744&\ -0.62\ \ \ 0.607&0.123 \\ \hline
  \end{tabular}\\
\end{center}
  \end{table}
The results of the calculations performed with different $g_A$ values are also presented,  in order to see which is the uncertainty associated with the use of the quenched (1.0) or unquenched (1.25, 1.275) values, in the $0\nu\beta\beta$ decay calculations.  
The GT components where HOC effects are partially included (without the tensor component), as expected, bring the main contribution to $M^{0\nu}_{total}$. The Fermi components are smaller than the GT components by a factor of 4-5 and are also diminished by the unquenched $g_A$ factors. The HOC do not affect the Fermi component of the $M^{0\nu}_{total}$, hence, their effects influence only the GT and tensor components. One can see that the tensor contributions are in the range of $4-9\%$ in all cases, and this justifies to this extent, its neglecting in some of the calculations from literature. 
We also take this opportunity to clarify the issue on the sign of the tensor term $M^{0\nu}_T$. Its effective sign  must be opposite to the GT one, as can be seen from table \ref{tab-ga} and (\ref{nme}), such that its inclusion contributes to the decrease of the total value of the $ M^{0 \nu}$. 
We also mention that the GT and F components have the same effective sign, hence their contributions add to $M^{0\nu}_{total}$.  Besides the errors associated with the use of different approximations when taking into account of the nuclear structure effects discussed above, there are other input parameters that can influence the NMEs values. One example is the axial-vector coupling constant, $g_A$. We mention that the majority of calculations in literature are performed using its unquenched value: 1.25. However, the most recent measurement reported a value a bit larger: 1.275 \cite{UCNA10}. In table 2 we present the calculations performed using both the quenched value, 1.0, and the two unquenched values, 1.25 and 1.275. 
One observes that differences between the quenched and unquenched results are within $(10-14)\%$, while differences between the older (1.25) and recent (1.275) unquenched values are negligible.  
This means that the issue of using a quenched or an unquenched value for $g_A$ in calculations, in connection to the nuclear structure effects, should be clarified, in order to eliminate the uncertainty associated to this choice, which is not negligible. Moreover, it is  important to mention, that this uncertainty does not refer to the additional $g_A^4$ multiplication factor that appears in the lifetimes formula, and which is usually considered separately. 
The differences in the lifetimes values associated only to this factor, amount to factors of 2.47 or 2.64, if one uses the unquenched values: 1.254 or 1.275, instead of the quenched one: 1.0.  

We also estimate that the error coming from the use of the value of $r_0 = 1.1 fm$ instead of $1.2 fm$ for this input parameter that appears in the nuclear radius formula, amounts to $\sim 7\%$. Also, in some calculations, different values of the cutoff parameters, that appear in the nuclear form factors, are used: 710 MeV instead of 850 MeV for $\Lambda_V$ and 850 MeV instead of 1086 MeV for $\Lambda_A$. Performing the calculations with (710, 850) MeV pair of values for the form factors, we got differences up to 8\% as compared with our results obtained with $\Lambda_V$ = 850 MeV, $\Lambda_A$ = 1086 MeV. 
Finally, we perform calculations, within the same approximation, with $\left< E\right>$ either fixed to 10 MeV (a value often used in calculations) or derived from the formula: $\left< E \right>=1.12 \ A^{1/2}$MeV, and we got the largest differences between results in the case of $^{48}Ca$, as large as $\sim 2\%$, that means, the results are practically insensitive to the change of $<E>$.

A general conclusion that comes out is that the inclusion of different nuclear effects and use of different input parameters such as those discussed above are important. Their cumulative effect may reflect in significant differences between the NMEs values found in literature, and hence a detailed knowledge of the errors associated with their inclusion/neglecting is needed in order to adequately compare to each other.  

In tables \ref{tab-ca}, \ref{tab-ge} and \ref{tab-se} we show the comparison between our $M^{0\nu}$ values and other results from the literature. To make the comparison more relevant, we compare as much as possible results obtained in the same approximation. We indicate in parenthesis the references which we compare to. We include the results similar to ours, obtained with different nuclear methods ShM, QRPA, IBA2 and EDS, and, where possible, provided by different groups. 
\begin{table}[tbh!]
\begin{center}
\caption[other48ca]{. Comparison of our $M^{0\nu}$ values with similar results for  $^{48}Ca$. \\
 H - HOC, F - FNS, J, A, C - Jastrow, AV-18 SRC, CD-BONN SRC, \\ U - UCOM SRC, $^*$ - KB3 effective interaction}.
\begin{tabular}{@{}l@{} l l l p{2.3 cm}@{}}
\hline
 Approach \ &H+F&H+F+J&H+F+A&H+F+C \\ \hline
 LSM&0.71\cite{HS10} 0.92$^*$\cite{npa818}&0.57\cite{HS10} 0.61$^*$\cite{npa818}&0.78\cite{HS10} 0.82$^*$\cite{npa818}&0.84\cite{HS10} 0.85(U)$^*$\cite{npa818} \\

 IBM-2& &1.98\cite{iba-prc2013}&2.28\cite{iba-prc2013}&2.38\cite{iba-prc2013}  \\

 EDF& & & &2.37(U)\cite{RMP10}  \\ \hline
 This paper \ &-0.77 0.93$^*$&-0.51 0.63$^*$&-0.73 0.89$^*$&-0.81 0.97$^*$ \\ \hline
\label{tab-ca}
 \end{tabular}
\end{center}
\end{table}

\begin{table}[tbh!]
\begin{center}
\caption[other76ge]{. Comparison of our $M^{0\nu}$ values with similar results for $^{76}Ge$. \\ H - HOC, F - FNS, J, A, C - Jastrow, AV-18 SRC, CD-BONN SRC \\
  U - UCOM SRC, $^*$ - GCN28.50 effective interaction }
\begin{tabular}{@{}l@{}p{1.55 cm}@{}p{1.55 cm}@{}p{1.55 cm}@{}p{1.55 cm}@{} p{1.55 cm}@{} p{1.55 cm}@{} p{1.55 cm}@{} p{1.55 cm}@{}}
\hline
 Approach &BARE&J&H&F&H+F&H+F+J&H+F+A&H+F+C \\ \hline
 LSM&4.04$^*$\cite{npa818}&2.85$^*$\cite{npa818}&3.29$^*$\cite{npa818}&3.45$^*$\cite{npa818}&2.96$^*$\cite{npa818}&2.30$^*$\cite{npa818}& &2.81$^*$(U)\cite{npa818} \\

 IBM-2& & & & & &5.42\cite{iba-prc2013}&5.98\cite{iba-prc2013}&6.16\cite{iba-prc2013}  \\

 EDF& & & & & & & &4.60(U)\cite{RMP10}  \\
 QRPA&8.53\cite{su-prc75-2007} 7.39\cite{sim-09}&4.46\cite{sim-09}&7.72\cite{su-prc75-2007}&7.03\cite{tu99} 6.14\cite{sim-09}&6.36\cite{su-prc75-2007} 5.63\cite{tu99}& -4.03\cite{su-jpsc173-2009} 4.72\cite{su-prc75-2007} 4.54\cite{sim-09}&5.91\cite{sim-09}&5.36(U)\cite{su-jpsc173-2009} 6.08(U)\cite{su-prc75-2007} \\  \hline
 This paper&\ 4.17&\ 3.03&\ 3.50&\ 3.62&\ 3.07&\ 2.38&\ 2.98&\ 3.18 \\ \hline
\label{tab-ge}
 \end{tabular} 
\end{center}
\end{table}

\begin{table}[tbh!]
\begin{center}
\caption[other82se]{. Comparison of our $M^{0\nu}$ values with similar results for $^{82}Se$.\\ H - HOC, F - FNS, J, A, C - Jastrow, AV-18 SRC, CD-BONN SRC \\
  U - UCOM SRC, $^*$ - GCN28.50 effective interaction}.
\begin{tabular}{@{}l@{} l l l l@{}}
  \hline
 Approach \ &H+F&H+F+J&H+F+A&H+F+C \\ \hline
 LSM&2.79$^*$\cite{npa818}&2.18$^*$\cite{npa818}& &2.64$^*$(U)\cite{npa818} \\

 IBM-2& &4.37\cite{iba-prc2013}&4.84\cite{iba-prc2013}&4.99\cite{iba-prc2013}  \\
 EDF& & & &4.22(U)\cite{RMP10}  \\
 QRPA&-2.91\cite{su-prc75-2007}&-2.77\cite{su-jpsc173-2009},\cite{su-prc75-2007}& &-3.72\cite{su-jpsc173-2009},\cite{su-prc75-2007}  \\ \hline
 This paper \ &-2.78&-2.18&-2.70&-2.88 \\ \hline
\label{tab-se}
 \end{tabular}
\end{center}
\end{table}
In general, one can see a very good agreement between our results and other ones obtained with ShM-based codes. Thus, the differences between similar results performed with ISM, reported in \cite{Cau95}-\cite{npa818} 
and ours are within a few percent. The comparison with results obtained with different methods is within the same limit of error, as has already been discussed in literature. 
For instance, the differences between our results and the results obtained with QRPA, IBM-2 and EDF are of about a factor of two. In case of $^{48}Ca$ one observes even a larger difference between ShM calculations, and IBA-2 and EDF. 
Particularly, for this isotope the ShM-based approaches are suited, since one can use in calculation the full pf valence shell as model space and a tested NN effective interaction. 
Indeed, the previous calculations of the NMEs for $2\nu\beta\beta$ decay mode from \cite{Cau90} predicted correctly the lifetime for this decay mode, before its measurement \cite{Bal96}. Further, it is worth to mention that in the case of $^{82}Se$, when the s.p. occupancies in the QRPA calculations are adjusted to the experimental ones \cite{Sch08}, NMEs values obtained with QRPA and ShM calculations get significantly closer \cite{su-jpsc173-2009}, within 30\%. In this situation, the treatment of uncertainties induced by the use of different nuclear effects discussed in this work becomes even more important, in the attempt to bring closer the NMEs values obtained with different nuclear methods.    

\section{Conclusions}
We analyze the effects of different nuclear ingredients such as SRC, FNS, HOC and nuclear input parameters such as $g_A$, $r_0$ and $<E>$, on the NMEs for $0\nu\beta\beta$ decay. 
The NMEs values are obtained with a new ShM code which allows a faster computation of the TBMEs. Their computation normally requires the numerical evaluation of two-dimensional integrals, one over the coordinate space and, the other, over the momentum space. In the actual version of the code we include the tensor component in the expression of the neutrino potential.

We study the effects of the nuclear ingredient  by gradually including them in calculations. Their common effect is to decrease the value of the bare $M^{0\nu}$. The SRC included by ``softer'' prescriptions than J-MS lead to a significant increase of the calculated $M^{0\nu}_{total}$ in literature, which amounts up to $\sim 30$\%.   
We found that the inclusion of only SRC by J-MS produces almost the same effect on the $M^{0\nu}_b$ value as that when all nuclear effects are included, but with SRC introduced by ``softer'' prescriptions, giving NMEs values which are close to the $M^{0\nu}_{total}$ ones within 4\% for the CCM-AV18 and 10\% for CCM-CDBonn parameterizations. 
This agreement is a rather accidental one, but is just an example of possible equivalent effects of some of these nuclear ingredients introduced in calculations.
The inclusion of HOC is important, the neglecting of these corrections results in larger values of $M^{0\nu}_{total}$ with up to $\sim 20\%$. However, the contribution of the tensor component is of 4-9\%, and justifies to this extent its neglecting in some calculations. We also mention that the effective sign of the tensor term $M^{0\nu}_T$ must be opposite to the GT one, as can be seen from table 2 and (\ref{nme}). 
Its inclusion contributes to the decrease of the total value of the $ M^{0 \nu}$, while the GT and F components give an additive contribution. 
We also found  that the use of different NN interactions can induce differences of the order of $\sim 17\%$ between NMEs values obtained within the same approximation.

Further, we analyze the errors associated with the use of different values of some nuclear input parameters.
The differences between the $M^{0\nu}$ values when using a quenched or unquenched value for the axial coupling constant $g_A$ are significant, i.e. within $(10-14)\%$, while differences between the older (1.25) and recent (1.275) unquenched values are negligible.  
This means that the issue of using a quenched or an unquenched value for $g_A$ in calculations, in connection to the nuclear structure effects, should be clarified, in order to eliminate the uncertainty associated to this choice. 
The use of different values of $r_0$, the constant that appears in the nuclear radius formula (1.1 fm instead of 1.2 fm) give an error of $\sim 7\%$, while the use of different values of the cutoff parameters that appear in the nuclear form factors, give an error of $\sim$ 8\%. 
Also, we find that the values of $M^{0\nu}$ values are not  sensitive to the change of the average energy $\left< E\right>$, used in the closer approximation.  
We conclude that the cumulative effect of all these nuclear effects is significant, and hence a convergence on the way of using them in the computation of the NMEs for $0\nu\beta\beta$ decay is needed in order to enter in an era of precision with these calculations. 
Further, we compare our results with other ones from literature. For ShM-based calculations the agreement is very good (within a few percent), while the differences with the results obtained with other nuclear structure methods are within the limits that have been discussed in literature, i.e. a factor of two. 
However, when the s.p. occupancies in the QRPA calculations are ajusted to the experimental ones, the NMEs values calculated with QRPA and ShM calculations get close within $\sim 30\%$. 
Thus, an appropriate treatment of errors induced by the different nuclear effects and input parameters becomes even more important in the attempt to get closer values of the NMEs for the $0\nu\beta\beta$ decay, when they are calculated with different nuclear methods. Also, a convergent view point on their inclusion and an uniformization of the calculations are needed as well, in order to enter in an era of precision concerning these calculations.  

\appendix
\section{Annex}
In this section we give a more detailed description of the numerical procedure which leads to the reduction of the complexity of computation of the neutrino potentials to one numerical integration over neutrino momentum.
We use the HO radial wave functions which can be expressed in terms of Laguerre associated polynomials:
\begin{equation}
\psi _{nl}(r) = N_{nl} \ exp \left(- \frac{\nu r^2}{2} \right) r^l \ L_{n}^{ \left( l+ \frac{1}{2} \right) }\nu r^2 \ ,
\end{equation}
where $\nu$ is the oscillator constant, $N_{nl}$ is the normalization constant
\begin{equation}
N_{nl} = \left[ \frac{2^n n!}{\left(2l +2n +1 \right)!!} \right]^{\frac{1}{2}} \left(2\nu \right)^{\frac{2l+3}{4}} \left( \frac{2}{\pi}\right) ^\frac{1}{4}
\end{equation}
and $L_{n}^{ \left( l+ \frac{1}{2} \right) }(\nu r^2)$ is the Laguerre associated polynomials:
\begin{equation}
L_{n}^{ \left( l+ \frac{1}{2} \right) }(\nu r^2) = \frac{\left(2l +2n +1 \right)!!}{2^n n!} \times \sum_{k=0}^{n}{n \choose {k}} \frac{1}{\left(2l +2k +1 \right)!!}\left( -2\nu r^2 \right)^k \ .
\end{equation}
In the expressions of the integrals over the neutrino potentials, the dependence on $r$ appears from the product of the $HO$ wave functions, the correlation function and Bessel functions. 
First, one can write the product of two $HO$ wave functions as a sum over the terms with the same power in $r$:
\begin{equation}
\fl \psi_{nl}(r) \psi_{n^\prime l^\prime}(r) = \sum_{s=0}^{n+n^{\prime}} A_{l+l^{\prime}+2s}(nl,n^{\prime}l^{\prime}) \left( \frac{2}{\pi} \right)^{\frac{1}{2}} \times (2 \nu )^\frac{l+l^{\prime}+2s+3}{2} e^{- \nu r^2} r^{l+l^{\prime}+2s},
\end{equation}
where $A_{l+l^{\prime}+2s}$ are coefficients independent of $r$ whose expressions are given in Eq.(\ref{alls}).
Then, one adds the contribution of the factor $\left[ 1+f(r) \right]^2$ that brings a dependence on $r$ in powers of $0$, $2$ and $4$:
\begin{equation}
\fl \left[ 1+f(r) \right]^2 = 1 -2 c e^{-ar^2} +2 b ce^{-ar^2}r^2 + c^2e^{-2ar^2} - 2 bc^2 e^{-2ar^2}r^2 + b^2 c^2e^{-2ar^2}r^4 \ .
\end{equation}
Finally, the computation of  the radial matrix elements requires to compute integrals of the form:
\begin{equation}
\nonumber \mathcal{I}_\alpha(\mu;m)  = \int_0 ^\infty q^2 dq \ V_\alpha(q) \times \left( \frac{2}{\pi} \right)^{\frac{1}{2}} \left( 2 \nu \right) ^{\frac{m+1}{2}} \int_0 ^\infty dr \ e^{-\mu r^2}r^m j_0(qr) \ ,
\end{equation}
where $\mu$ = $\nu$, $\nu+a$, $\nu+2a$ and  $m$ is integer. In this expression the integration over $r$ can be done analytically and one gets:
\begin{eqnarray}
\nonumber \left( \frac{2}{\pi} \right)^{\frac{1}{2}} \left( 2 \nu \right) ^{\frac{m+1}{2}} \int_0 ^\infty dr \ e^{-\mu r^2}r^m j_0(qr)  = \\
 \left( \frac{2\nu}{2\mu} \right) ^{\frac{m+1}{2}} \times \left( m-1 \right)!!  \sum_{k=0}^{\frac{m}{2}-1} (-1)^k {{\frac{m}{2}-1} \choose {k}} \frac{e^{-\frac{q^2}{4\mu}} }{(2k+1)!!(2\mu)^k}q^{2k} \ .
\end{eqnarray}
Thus, $\mathcal{I}_{\alpha}(\mu;m)$ becomes:
\begin{equation}
\nonumber \mathcal{I}_{\alpha}(\mu;m) = \left( \frac{2 \nu}{2\mu} \right) ^{\frac{m+1}{2}} (m-1)!! \times \sum_{k=0}^{\frac{m}{2}-1}(-1)^{k} {\frac{m}{2}-1 \choose {k}} \mathcal{J}_{\alpha}(\mu;k) \ ,
\end{equation}
where $\mathcal{J}_{\alpha}(\mu;k)$ are integrals over  momentum:
\begin{equation}
\mathcal{J}_{\alpha}(\mu;k)  = \frac{1}{(2k+1)!!} \frac{1}{(2\mu)^k} \times \int_0 ^\infty exp \left(- \frac{q^2}{4\mu} \right) q^{2k +2}V_{\alpha}(q) G^2_{\alpha}(q^2)dq
\label{qintegral}
\end{equation}
The expression for $A_{l+l^{\prime}+2s}(nl,n^{\prime} l^{\prime})$ used in (\ref{finaleq}) is:
\begin{eqnarray}
\nonumber A_{l+l^{\prime}+2s}(nl,n^{\prime} l^{\prime}) = \left[ \frac{n!(2l+2n+1)!!}{2^n} \frac{n^{\prime}!(2l^{\prime}+2n^{\prime}+1)!!}{2^{n^{\prime}}} \right] \\
 \times  (-1)^s \sum_k \frac{1}{k! (n-k)!(2l+2k+1)!!} \frac{1}{k^\prime ! (n^\prime-k^\prime)!(2l^\prime+2k^\prime+1)!!} \ ,
\label{alls}
 \end{eqnarray}
with  $max(0,s-n^{\prime}) \leq k \leq min(n,s)\ , \ \ \ k + k^\prime = s$ \ . \\
The expression for $\mathcal{K}_{\alpha}(m)$ used in (\ref{finaleq}) is:
\begin{eqnarray}
\fl \nonumber \mathcal{K}_{\alpha}(m)  =  \frac{1}{\sqrt{2\nu}} [ \mathcal{I}_{\alpha}(\nu ; m) - 2c\mathcal{I}_{\alpha}(\nu + a;m) + 2c \left( \frac{b}{2\nu} \right) \mathcal{I}_{\alpha}(\nu +a;m+2) \\
\fl + c^2\mathcal{I}_{\alpha}(\nu +2a;m)-2c^2 \left( \frac{b}{2\nu} \right) \mathcal{I}_{\alpha}(\nu +2a;m+2) + c^2\left( \frac{b}{2\nu} \right)^2 \mathcal{I}_{\alpha}(\nu +2a;m+4) ]  \ ,
\label{kalpha}
\end{eqnarray}
where $a$, $b$, and $c$ are the SRC parameters entering (\ref{src}).
\section*{Acknowledgments}
This work was done with the support of the MEN and UEFISCDI through the project IDEI-PCE-3-1318, contract Nr. 58/28.10/2011.


\section*{References}
\begin{thebibliography}{99}
\bibitem{AEE08} F. T. Avignon, S.R. Elliott and J. Engel, Rev. Mod. Phys. {\bf 80}, 481 (2008).
\bibitem{EJI10} H. Ejiri. Prog. Part. Nucl. Phys., {\bf 4}, 249 (2010).
\bibitem{VES12} J. Vergados, H. Ejiri and F. Simkovic, Rep. Prog. Phys. {\bf 75}, 106301 (2012).
\bibitem{engh_jpg2012} J. Engel, J.Phys. G: Nucl. Part. Phys. {\bf 39}, 124001  (2012).
\bibitem{VOG12} P. Vogel, J.Phys. G: Nucl. Part. Phys. {\bf 39}, 124002  (2012).
\bibitem{su_jpg2012} J. Suhonen and O. Civitarese, J.Phys. G: Nucl. Part. Phys. {\bf 39}, 124005  (2012).
\bibitem{FAE12} A. Faessler, V. Rodin, F. Simkovic, J.Phys. G {\bf 39}, 124006 (2012).
\bibitem{ROJ12} W. Rodejohann, J.Phys. G: Nucl. Part. Phys. {\bf 39}, 124008 (2012).
\bibitem{ROD07} V.A. Rodin, A. Faessler, F. Simkovic and P. Vogel, Phys. Rev. C {\bf 68}, 044302 (2003); Nucl. Phys. A {\bf 766}, 107 (2006); Nucl. Phys. A {\bf 793}, 213 (2007).
\bibitem{KOR07} M. Kortelainen, O. Civitarese, J. Suhonen and J. Toivanen, Phys. Lett. B {\bf 647}, 128 (2007); Phys. Rev. C {\bf 76}, 024315 (2007).
\bibitem{SIM08} F. Simkovic, A. Faessler, V.A. Rodin, P. Vogel and J. Engel, Phys. Rev. C {\bf 77}, 045503 (2008).
\bibitem{sim-09} F. Simkovic, A. Faessler, H. Muther, V. Rodin, and M. Stauf, Phys. Rev. C {\bf 79}, 055501 (2009).
\bibitem{SK01}S. Stoica and H.V. Klapdor-Kleingrothaus, Nucl. Phys. A {\bf 694}, 269 (2001).
\bibitem{Cau95} E. Caurier, A.P. Zuker, A. Poves, G. Martinez-Pinedo, Phys. Rev. C {\bf 50}, 225 (1994); J. Retamosa, E. Caurier and F. Nowacki, Phys. Rev. C {\bf 51}, 371 (1995)
\bibitem{prl100} E. Caurier, J. Menendez, F. Nowacki, and A. Poves, Phys. Rev. Lett. {\bf 100}, 052503 (2008).
\bibitem{npa818} J. Menendez, A. Poves, E. Caurier, F. Nowacki, Nuclear Physics {\bf A 818}, 139–151 (2009).
\bibitem{HS10} M. Horoi and S. Stoica, Phys. Rev. {\bf 81}, 024321 (2010).
\bibitem{BI09} J. Barea and F. Iachello, Phys. Rev. C {\bf 79}, 044301 (2009).
\bibitem{iba-prl12} J. Barea, J. Kotila, and F. Iachello, Phys. Rev. Lett. {\bf 109}, 042501 (2012).
\bibitem{iba-prc2013} J. Barea, J. Kotila, F. Iachello, Phys. Rev. C {\bf 87}, 014315 (2013).
\bibitem{RAH10} P.K. Rath, R. Chandra, K. Chaturvedi, P.K. Raina, J.G. Hirsch, Phys. Rev. C {\bf 82}, 064310 (2010).
\bibitem{RMP10} T.R. Rodriguez and G. Martinez-Pinedo, Phys. Rev. Lett {\bf 105}, 252503 (2010).
\bibitem{NSH12} A. Neacsu, S. Stoica and M. Horoi, Phys, Rev. C {\bf 86}, 067304 (2012).
\bibitem{EH09} J. Engel and G. Hagen, Phys. Rev. C {\bf 79}, 064317 (2009); J. Engel, AIP Conf. Proc. (MEDEX-11) {\bf 1417}, 42 (2011); J.D. Holt and J. Engel, Pys. Rev. C {\bf 87}, 064315 (2013).
\bibitem{medex11-mh} M. Horoi, AIP Conf. Proc. (MEDEX-11) {\bf 1417}, 57 (2011).
\bibitem{jastrow55} R. Jastrow Phys. Rev. {\bf 98}, 1479 (1955).
\bibitem{jastrow} G. A. Miller and J. E. Spencer, Ann. Phys. {\bf 100}, 562 (1976).
\bibitem{ccm} C. Giusti, H. Muther, F. D. Pacati, and M. Stauf, Phys. Rev. C {\bf 60}, 054608 (1999).
\bibitem{ccm1} H. Muther and A Polls, Phys. Rev. C 61, 014304 (1999) Part. Nucl. Phys. {\bf 45}, 243 (2000).
\bibitem{ANT} E. Caurier, code Antoine, unpublished.
\bibitem{gxpf1a}M. Honma, T. Otsuka, B.A. Brown and T. Mizusaki, Phys. Rev. C {\bf 69}, 034335 (2004);  M. Honma, T. Otsuka, B. A. Brown, and T. Mizusaki, Eur. Phys. J. A {\bf 25}, Suppl. 1, 499 (2005)
\bibitem{KB68} T.T.S. Kuo and G.E. Brown, Nucl. Phys. A {\bf 114}, 235 (1968).
\bibitem{jun45} M. Honma, T. Otsuka, T. Mizusaki and M. Hjorth-Jensen, Phys. Rev. C {\bf 80}, 064323 (2009).
\bibitem{feldmeier-ucom}  H. Feldmeier, T. Neff, R. Roth, J. Schnack, Nucl. Phys. A {\bf 632}, 61 (1998).
\bibitem{suhonen-ucom} M. Kortelainen, O. Civitarese, J. Suhonen, J. Toivanen, Physics Letters B {\bf 647}, 128–132 (2007).
\bibitem{su-prc75-2007} Markus Kortelainen and Jouni Suhonen, Phys. Rev. C {\bf 75}, 051303(R) (2007).
\bibitem{tu99} F. Simkovic, G.Pantis, J.D. Vergados, A. Faessler, Phys. Rev. C {\bf 60}, 055502 (1999).
\bibitem{UCNA10} UCNA Collaboration: J. Liu et al., Phys. Rev. Lett. {\bf 105},181803 (2010).
\bibitem{Cau90} E. Caurier, A. Poves and A.P. Zuker, Phys. Lett. B {\bf 252}, 13 (1990).
\bibitem{Bal96} A. Balysh et al., Phys. Rev. Lett. {\bf 77}, 5186 (1996).
\bibitem{Sch08} J. Schiffer et al., Phys. Rev. Lett. {\bf 100}, 112501 (2008); B.P. Kay, et al., Phys. Rev. C {\bf 79}, 021301 (R) (2009).
\bibitem{su-jpsc173-2009} Osvaldo Civitarese and Jouni Suhonen, Journal of Physics: Conference Series {\bf 173}, 012012 (2009).
\end{thebibliography}
\end{document}